# (Super)Spreading and Drying of Trisiloxane-Laden Quantum Dot Nanofluids on Hydrophobic Surfaces

Nikolai Kubochkin[1*], Joachim Venzmer[2], Tatiana Gambaryan-Roisman[1]

[1]Institute for Technical Thermodynamics, Technische Universität Darmstadt, Alarich Weiss Strasse 10, 64287 Darmstadt, Germany
[2]Research Interfacial Technology, Evonik Nutrition & Care GmbH, Goldschmidtstr. 100, 45127 Essen, Germany

*corresponding author: kubochkin@ttd.tu-darmstadt.de

**Abstract:** Nanofluids hold promise for a wide range of areas of industry. However, understanding of wetting behavior and deposition formation in course of drying and spreading of nanofluids, particularly containing surfactants, is still poor. In this paper, the evaporation dynamics of quantum dot-based nanofluids and evaporation-driven self-assembly in nanocolloidal suspensions on hexamethyldisilazane-, polystyrene-, and polypropylene-coated hydrophobic surfaces have been studied experimentally. Moreover, for the very first time, we make a step to understanding of wetting dynamics of superspreader surfactant-laden nanofluids. It was revealed that drying of surfactant-free quantum dot nanofluids in contrast to pure liquids undergoes not three but four evaporation modes including last additional pinning mode when contact angle decreases whilst triple contact line is pinned by the nanocrystals. In contrast to previous studies, it was found out that addition of nanoparticles to aqueous surfactant solutions leads to deterioration of spreading rate and to formation of double coffee ring. For all surfaces examined, superspreading in presence and absence of quantum dot nanoparticles takes place. Despite the formation of coffee rings on all substrates, they have different morphology. Particularly, the knot-like structures are incorporated into the ring on hexamethyldisilazane- and polystyrene-coated surfaces.

## 1. Introduction

Liquid systems containing suspended colloids like particles, micelles, vesicles, proteins are omnipresent in everyday life and nature. One can observe the result of interaction of such complex liquids with solid surfaces occasionally spilling coffee over a table or water over a dusty surface, washing dishes with use of detergent or when cooking. Applications of complex liquids laden with colloids hold promise for a wide range of areas of industry including ink-jet printing, optics and lighting industry, agriculture, cosmetics, and oil recovery.[1-4] Additionally, good understanding of colloidal suspensions behavior at different conditions can contribute significantly to medicine, biotechnology development, food industry, and forensics, since all biological liquids are ingenious compositions of base liquids and different types of colloids.[5-7] In recent years, a novel type of complex liquids containing nanoparticles that is commonly referred to in literature as nanofluids has drawn the attention of researchers due to their unique thermal and wetting properties dramatically distinguishing from the properties of pure liquids.[8-10] One of the main topics research focuses on is the influence of different factors on the deposit formation after drying of nanofluids.[11-12]

As revealed in a pioneering paper,[13] in the absence of any surface-active additives the typical ring-shaped deposit – coffee ring – is formed after drying of a sessile colloidal droplet. This effect is engendered by the non-uniform evaporation flux over the droplet surface actuating the internal outward capillary flows within the droplet tending to replenish the mass loss from the droplet periphery. Evaporating droplet stays pinned and losses its volume via the contact angle diminishing. Nevertheless, even in at first glance relatively simple case of drying of sessile droplet not only coffee ring but also a manifold of different types of deposits can be obtained depending on the type of particles, base liquids and substrates chosen, and surrounding conditions. That can be explained by the fact that besides capillary flows, circulating Marangoni flows induced by the surface tension differences over the droplet surface can essentially affect the final deposition pattern and can cause uniform distribution of particles within the wetted perimeter. Marangoni flows, either thermocapillary or solutocapillary, can result from decreasing temperature on the droplet edge caused by evaporation or nanoparticle-induced surface tension alteration in close vicinity to the droplet edge, accordingly.[14-15] The data on influence of nanoparticles on the base fluid surface tension for a wide range of nanoparticle type is summarized in the review by Estelle et.al.[16]

Flows within a droplet are believed to be only one of multiple keys to understanding of drying dynamics and deposition phenomena. Different hypotheses were developed to explain the formation of different deposition patterns in course of drying and transitions between them.[14,17] Some of works aim to investigate the role of surface (intermolecular) forces in the deposit formation. The transition from ring-shaped to uniform pattern was explained by the interplay of hydrodynamic and van der Waals forces by Sommer et.al.[18] Bhardwaj et.al. employed DLVO theory of colloidal stability (named after Derjaguin, Landau, Verwey, and Overbeek) to show that the uniform particle coverage of wetted region is caused by the DLVO attraction forces between particles and substrate.[19]

Suspended colloids can have a tremendous impact on the deposit formation not only via droplet drying but also via droplet spreading. Wasan et.al. were the first who revealed that nanoparticles being added to the base fluid can enhance its spreading rate.[10] The acceleration of spreading of liquid droplets containing nanoparticles compared to droplets of base liquids has been confirmed by further investigation.[3,20-24]

Some authors, when considering wetting dynamics, emphasize the role of disjoining pressure (or so-called Derjaguin's pressure) acting in a thin film formed in the close vicinity to the triple contact line (TCL).[25-30] The disjoining pressure is generally presented as a sum of three components. The first two of them are referred to as van der Waals (molecular) component and electrostatic component, respectively, and can be estimated in the frame of DLVO theory.[25-29] The third, structural, component of disjoining pressure is of entropic nature and arises from changes of the dynamic structure of liquids in thin films confined by solid-liquid and liquid-vapor interfaces, respectively.[26,29,30] If colloids are added to a base pure liquid, interesting effects mainly governed by the aforementioned intermolecular forces acting between them can affect the form of disjoining pressure isotherm and hereby alter the wetting dynamics. Apparently, changes in the wetting dynamics lead to changes in the final drying patterns. While micro-sized particles in most cases cannot break into the droplet wedge and only hinder the advancing wetting front propagation by pinning of the TCL, the wetting behavior of nanocolloidal suspensions and physics behind is more complex.[31]

The peculiar trait of nanofluids is in tending of micelles, nanoparticles and other nano-sized colloids to self-assemble in the thin films and render an additional pressure in these spaces usually referred to as structural pressure.[21-23] It was reported that structural forces oscillate and decay with increase of the film thickness resulting in pressure gradient in the confined films, which, in turn, can cause enhanced spreading of droplet of nanofluid compared to its base fluid. Interesting results on the influence of a structural disjoining pressure and on a role of nanoparticle ordering in a precursor film adjoining to the droplet bulk were provided by Kondiparty et.al.[23] In particular, it was shown that the magnitude of the structural disjoining pressure is higher for small particles. Higher spreading velocities of nanofluids compared to pure liquids were also observed by Sefiane et. al.[24] Adsorbed nanoparticles, according to authors, can play a role of a buffer between base fluid and substrate and cause spreading enhancement of nanofluids reducing friction. It is of importance to underscore that in these works acceleration of spreading of liquids containing nanoparticles compared to base liquids was reported. In contrast, the deterioration of the droplet spreading rate for dilute nanofluids was found by Lu et.al.[32] Authors attribute slower contact line velocity to changes of surface tension and viscosity of nanofluids. Another example of deterioration of wetting was observed by Vafaei et.al.[33] It was reported that bismuth telluride nanoparticles added to pure water significantly change static contact angle on hydrophilic glass and silicon wafers. This effect, according to authors, can be related to repulsion forces between functional groups covering particles and hydroxyl groups on silicon and glass surfaces. As is in work by Kondiparty et.al.,[23] it was reported by Vafaei et.al.[33] that smaller nanoparticles bring more pronounced wetting alteration.

As is clear form the last example, additionally to structural, another surface force able to contribute to wetting dynamics and colloidal self-assembly is tightly related to the type of nanoparticles used. Some types of nanoparticles (chemically stabilized nanoparticles, quantum dots) possess a shell of ligands, attached to the core. This shell is to protect particles from core-core van der Waals interactions, which are known to be very strong in case, for instance, metallic nanoparticles. Moreover, the screening allows to maintain stability of suspension and to tune lattice structures.[34] Being a protection on the one hand, this shell existence, on the other hand, leads to the steric hindrance forces appearing when capping ligand brushes come close to each other in confinement. The latter can be exemplified by an elastic repulsion induced by the surface-bounded molecules chain compression.[35] Therefore, steric forces alongside with entropic forces can affect final deposits.

Note that a major part of works considers spreading of water-based nanofluids. However, most surfaces are poorly wetted by water and require addition of surface active agents to colloidal suspensions to reach better wetting performance.[36-39] Interactions of nanofluids with solids, however, involve much more complex physical and chemical processes on different length scales behind them if, besides nanoparticles, surfactants are added to a base fluid.[40] Despite a widespread interest and intensive investigations, the information on the surfactant-laden nanosuspension wetting behavior is still quite limited and inconclusive. To the best of authors' knowledge, a recent study on spreading dynamics of nanofluids containing surfactant presented by Harikrishnan et.al.[40] is still the only work comparing wetting behavior of surfactant-laden nanosuspensions with water-based nanosuspensions and particle-free surfactant solutions. It has been found out that in complex systems consisting of nanoparticles and surfactants, the latter dominate spreading dynamics. Conventional SDS and CTAB surfactants were used in their investigation.[40]

If the surface energy of solid is low, a major part of commonly used surfactants can only decrease contact angle but do not provide complete wetting of surface. It was shown, for instance, by Dutschk et.al.[41] that SDS, DTAB, and DTAS at critical aggregation concentration (CAC) spread until final finite contact angle is reached over surfaces with contact angle of water ~ 80° whilst do not spread at all over the surfaces with contact angle of water higher that 90°. Fast spreading over hydrophobic surfaces until reaching *zero* contact angle is referred to as superspreading effect and is generally associated with trisiloxane (silicone-based) surfactants.[2,39,42,43] According to Svitova et.al.,[44] trisiloxane surfactants are known to have two critical concentrations: CAC and CWC (critical wetting concentration), respectively. At $c_s$ < CWC, aqueous solutions show partial wetting of hydrophobic surfaces whereas at $c_s$ > CWC famous superspreading phenomenon occurs.[44,45]

Notwithstanding the fact that superspreading effect as well as processes of nanoparticles (particularly, quantum dots) self-assembly are under intense investigation in recent years,[46-49] they surprisingly have never been overlapped. In this work, for the very first time, we make a step to understanding of wetting dynamics of superspreader surfactant-laden quantum dots nanofluids on surfaces with different surface properties. Besides, to the best of authors' knowledge, this work is the first attempt to cover by an experimental investigation evaporation dynamics of quantum dot-based nanofluids.

This paper is organized as follows. We first introduce materials and methods used in the next section. In section 3, we present and discuss experimental results on drying dynamics of surfactant-free nanofluids and deposition patterns formed followed by the results on wetting dynamics and residual patterns of surfactant-laden nanofluids for two surfactant concentrations including superspreading concentration. After that, we draw conclusions in section 4.

## 2. Materials and Methods
*2.1. Surfaces Manufacturing*

Two types of surfaces were used in experiments: substrates manufactured by spin coating with hydrophobic polymers and substrates produced by low pressure chemical vapor deposition of silane.

The preparation of surfaces for spin coating procedure was conducted according to the following protocol: glass microscope slides (Marienfeld, Carl Roth, Germany) were cut into pieces 3.5 x 2.5 cm² and intensively rinsed with ultrapure water Milli-Q (18.2 MΩ·cm at 25 °C), etched in Piranha solution (sulfuric acid and hydrogen peroxide, 1:1) for 30 min to remove all organic residues, rinsed with ultrapure water again and dried in a strong nitrogen jet. Polystyrene (molecular weight $M_W$ = 35000) and polypropylene (molecular weight $M_W$ = 12000) were purchased in Sigma Aldrich (Germany) and were in form of pellets. Polystyrene- and polypropylene-coated substrates were produced according to the following protocol: 6wt% solution of polystyrene (it is further denoted as PS) in Toluene (Carl Roth, Germany) was prepared and left at room temperature (approximately 25°C) until complete polymer dissolution for one week. The solution was deposited with use of the spin coater (WS 400B 6NPP lite, Laurell Technologies) at 800 rpm for 30 s. After that, substrates were blown in a strong nitrogen jet and were ready to use. The use of toluene, however, was not favorable for dissolving of polypropylene (it is further denoted as PP) because its boiling temperature ($T_b$ = 110 °C*) is lower than the melting temperature of polypropylene ($T_m$ = 160 °C*); hence, decaline ($T_b$ = 190 °C*, Carl Roth, Germany) was used as an appropriate solvent. 2wt% solution of PP in decaline was prepared, heated up to 185°C and stirred at 700 rpm at 185°C for approximately four hours. The solution obtained was spin coated at 3000 rpm for 60 s. Note that the glass slides as well as the droppers were preliminary heated up to 100-120°C and kept at this temperature to provide better wettability of the surface and to avoid cracks in the course of a polymer film formation when evaporating. The spin coater vacuum chuck could not be heated and due to that the timespan between the removal of substrates, droppers and polymer solution from the hot plates and the moment the deposition began was minimized to 20 s. Freshly polypropylene-coated glass slides were baked on the hot plate for 5 min at (110±10)°C to let the residual solvent evaporate.

HMDS (1,1,1,3,3,3-Hexamethyldisilazane, Carl Roth, ≥98 %, for GC, Germany) was used as a silanizing agent to prepare a hydrophobic surface. The silanization of microscope glass slides was performed by low pressure chemical vapor deposition, according to the following protocol: slides (Marienfeld, Carl Roth, Germany) were cut into pieces 3.5 x 2.5 cm², carefully pre-rinsed in Milli-Q (18.2 MΩ·cm at 25 °C) to remove dirt and turbid spots, ultrasonificated in acetone (Carl Roth, Germany), ethanol (Carl Roth, Germany) for 15 min in each solvent, then ultrasonificated in Milli-Q ultrapure water for 15 min and dried in a strong nitrogen jet. Before silanization, surfaces of glass slides were activated with oxygen plasma (Diener electronic GmbH & Co KG, 50 W, 1 min). To silanize surfaces with HMDS, they were put into a desiccator with a vial filled with ~ 3 ml of HMDS. The desiccator was evacuated with a pump and left for 4 days to obtain better surface coverage. Freshly prepared substrates were dried in a strong nitrogen jet.

The topography of different areas (squares $160~\mu m \times 160~\mu m$) of substrates was scanned and investigated with Confocal Profilometer μSurf (NanoFocus Expert) (Fig. 1). The surface roughness of substrates was evaluated according to ISO-4287 and averaged for three different areas for each of three different substrates. The result is presented in Table 1.

The bubble-like structure and high values of roughness of PP-coated substrates are most probably related to non-uniform cooling down of the substrate, droppers and polymer solution before placing in onto the vacuum chuck of the spin coater.

The wettability of substrates was investigated with droplet shape analyzer Kruss DSA-100 by placing a small water droplet $(2.2 \pm 0.1)\mu l$ onto the modified glass substrate and measuring the static contact angle $\theta_w$ (Fig. 1). Higher surface roughness in case of PP implies higher apparent contact angle due to heterogeneous wetting: the intrinsic water contact angle on a smooth PP-covered surface is $\theta_w = 100°$, according to results provided by Lock et.al.[50]

---

* Data taken from MSDS

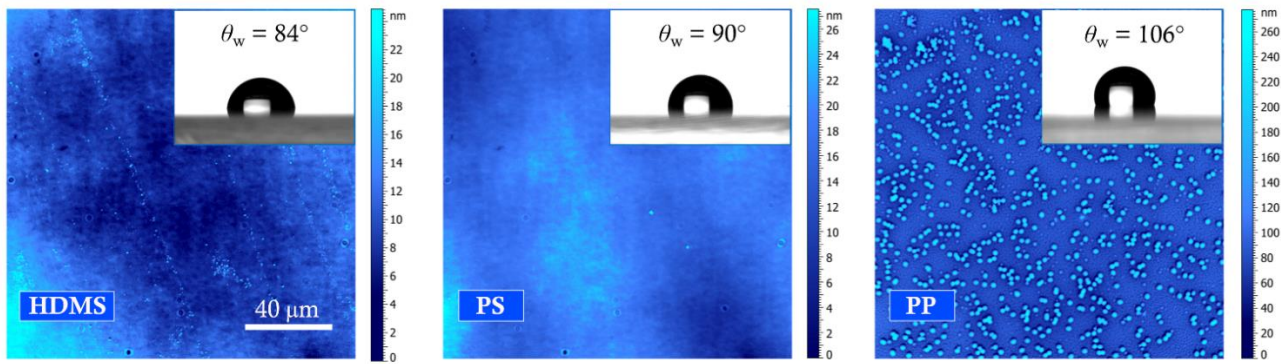

**Figure 1.** Topography of surfaces (squares 160 $\mu m$ × 160 $\mu m$) investigated with Confocal Profilometer µSurf (NanoFocus Expert), inner microphotographs illustrate static wetting by water droplet (2.2 $\mu l$). From left to right: HDMS ($\theta_w = 84°$), PS ($\theta_w = 90°$) and PP ($\theta_w = 106°$).

**Table 1.** Substrate Properties.

| Substrate  Property | PS | PP | HDMS |
|---|---|---|---|
|  | Spin coated | | Silanized |
| $\theta_w$, ° | 90±2 | 105±2 | 84±2 |
| $R_z$, nm  Maximum Height of roughness profile | 6.6±1.8 | 203.0±83.5 | 7.1±4.1 |
| $R_q$, nm  Root-mean-square (RMS) deviation of the roughness | 1.2±0.3 | 36.9±20.1 | 1.0±0.3 |
| $R_p$, nm  Maximum peak height of the roughness profile | 3.6±1.4 | 140.0±61.2 | 4.3±3.5 |

*2.2. Liquids and Nanoparticles Used*

Quantum dots (denoted as QDs hereafter) are semiconductor-based nanocrystals. They are also known as man-made "artificial atoms" due to their tailor-made properties. For our experiments, the hydrophilic water-soluble QD nanocrystals – Cadmium Telluride CdTe (emission maximum $\lambda = 650 \pm 5\ nm$) terminated with mercaptosuccinic acid (MSA) – were purchased from PlasmaChem GmbH (Germany). This type of the quantum dots is stable in the pH region of 5-11. The average core size of CdTe quantum dots usually does not exceed 10 nm. The core size of the quantum dots nanocrystals, i.e. the size of nanoparticles without taking into account the thickness of the ligand brush, varies depending on the emission maximum wavelength. According to Peng's equation,[48,51] mean diameter $d$ (in nm) of CdTe QDs can be estimated as follows

$$d = \sum_{i=0}^{3}(-1)^{i+1}\lambda^i c_i$$

where $\lambda$ (in nm) is the wavelength of the first excitonic absorption peak (in nm), $c_i$ are coefficients, $c_3 = 9.8127 \cdot 10^{-7}$, $c_2 = 1.7147 \cdot 10^{-3}$, $c_1 = 1.0064$, $c_0 = 194.84$. Assuming $\lambda \approx 610\ nm$ from the photoluminescence and UV-vis spectra of QDs*, one can obtain $d \approx 3.8\ nm$. The experiments have been performed at $c_{QD} = 0.5$ wt%.

---

* Data taken from MSDS

Polyalkyleneoxide modified heptamethyltrisiloxane, which is a nonionic silicone-based surfactant better known as Silwet L-77, was used in experiments. For the sake of simplicity, trisiloxane surfactants are named $T_m$, where m is a number of oxyethylene groups. As a commercial product, Silwet L-77 can be roughly considered as $T_{7.5}$ and is usually referred to in literature as a superspreader.[43] Silwet L-77 (Momentive, Germany) was used without any further purification.

The surfactant solutions were prepared with ultrapure water, rigorously shaken by hand and then ultrasonificated for 15 min (Elmasonic S10H, Elma Schmidbauer GmbH). Nanofluids were prepared by adding of prepared surfactant solution into a vial with the quantum dots powder to obtain a desired concentration and were ultrasonificated for 15 min. The dynamic surface tension for each surfactant solution as well as for water to ensure the purity was preliminary measured by pendant drop method and showed reproducibility. The surface tension of QDs nanofluid ($c_{QD} = 0.5\ wt\%$) in absence of surfactant was equal to $72.4 \pm 0.7\ mN/m$. Roques-Carmes et.al. performed surface tension measurements for a similar system – water suspension of core/shell CdTe/CdS QDs terminated with 3-Mercaptopropionic acid.[52] It has been noticed a slight decrease of surface tension (~ 2 mN/m), which means that QDs did not generally tend to adsorb at water/air interface. The surface tension of SilwetL-77 aqueous solution at CAC measured after 1 min is $\gamma_S = (23.9 \pm 0.5)\ mN/m$. All solutions and suspensions were used within 8 hours since the moment they were prepared to avoid hydrolysis. The surfactant concentrations used were 0.007 wt% (critical aggregation concentration, CAC)[53] and 0.1 wt% corresponding to optimum wetting performance of Silwet.[2] All experiments were performed at room temperature and relative humidity (T = (23±2) °C, RH = (30±5) %), unless otherwise stated.

*2.3. Experimental Methods and Techniques*

The sessile drop and the pendant drop experiments were conducted with use of droplet shape analyzer Krüss DSA-100 (Germany). Density of Silwet L-77 aqueous solution was considered to be the same as density of water due to small concentration of surfactant. Density of QD nanosuspension was calculated as follows $\rho_{ns} = \rho_w(1 - c_{QD}) + \rho_{QD} c_{QD}$, where $\rho_w = 997\ kg/m^3$ is density of water and $\rho_{QD} = 5850\ kg/m^3$ is density of cadmium telluride.[54]

In spreading experiments, droplets of $(2.2 \pm 0.1)$ µl volume were gently deposited onto the substrates with a gastight microsyringe (Hamilton, USA). The radius of the droplet is less than the capillary length, which allows to neglect gravity effect and use the spherical cap approximation for base diameter, contact angle and volume estimation. The side view camera inclination angle was set 2° to obtain the clearly distinguishable drop-reflection pair for the baseline setting. The brightness average level was set to obtain the contrast droplet contour but at the same time to minimize the contact angle determination errors due to underexposure and overexposure. Recorded videos (50 fps) were splitted and image sequences were processed with the home-made software to obtain droplet base diameter, contact angle and volume time dependencies.

The experiments on comparison of the particle-laden and particle-free surfactant solution spreading behavior were performed by using the same base surfactant solutions (i.e. nanoparticles were added to the surfactant solution prepared and examined before) to eliminate the possible errors related to surfactant solution preparation and behavior.

The presented results on the surfactant-free nanofluid droplet evaporation dynamics are not averaged data but the data obtained during one of experiments. This has been done since the moment when contact line started to recede in the mixed regime evaporation slightly varied from experiment to experiment and the sharp transition from constant contact angle regime to mixed regime and additional constant contact radius regime could be smoothed when averaging. Nevertheless, calculation of standard deviation for experimental data (for first two regimes) showed that it does not exceed 0.1 mm for droplet diameters and 5° for droplet contact angles which allows to conclude that each test represents the evaporation dynamics well. The experimental results on the surfactant-laden nanofluid droplet spreading dynamics are the averaged data of 3-8 measurements.

The deposits left on the substrates after droplet evaporation were investigated with Confocal Profilometer µSurf (Nanofocus AG, Germany) and post-processed with µSurf software.

## 3. Results and Discussions

### *3.1. Drying of Quantum Dots-Laden Nanofluid*

In this subsection, we study how evaporation dynamics of water droplets is affected by the presence of nanoparticles and substrate wettability. Goniometric investigations of evaporation process to obtain droplet base diameter, contact angle and volume evolution on HDMS, PS, and PP surfaces were performed with pure water and water containing hydrophilic water-soluble QDs. The results are shown in Fig. 2.

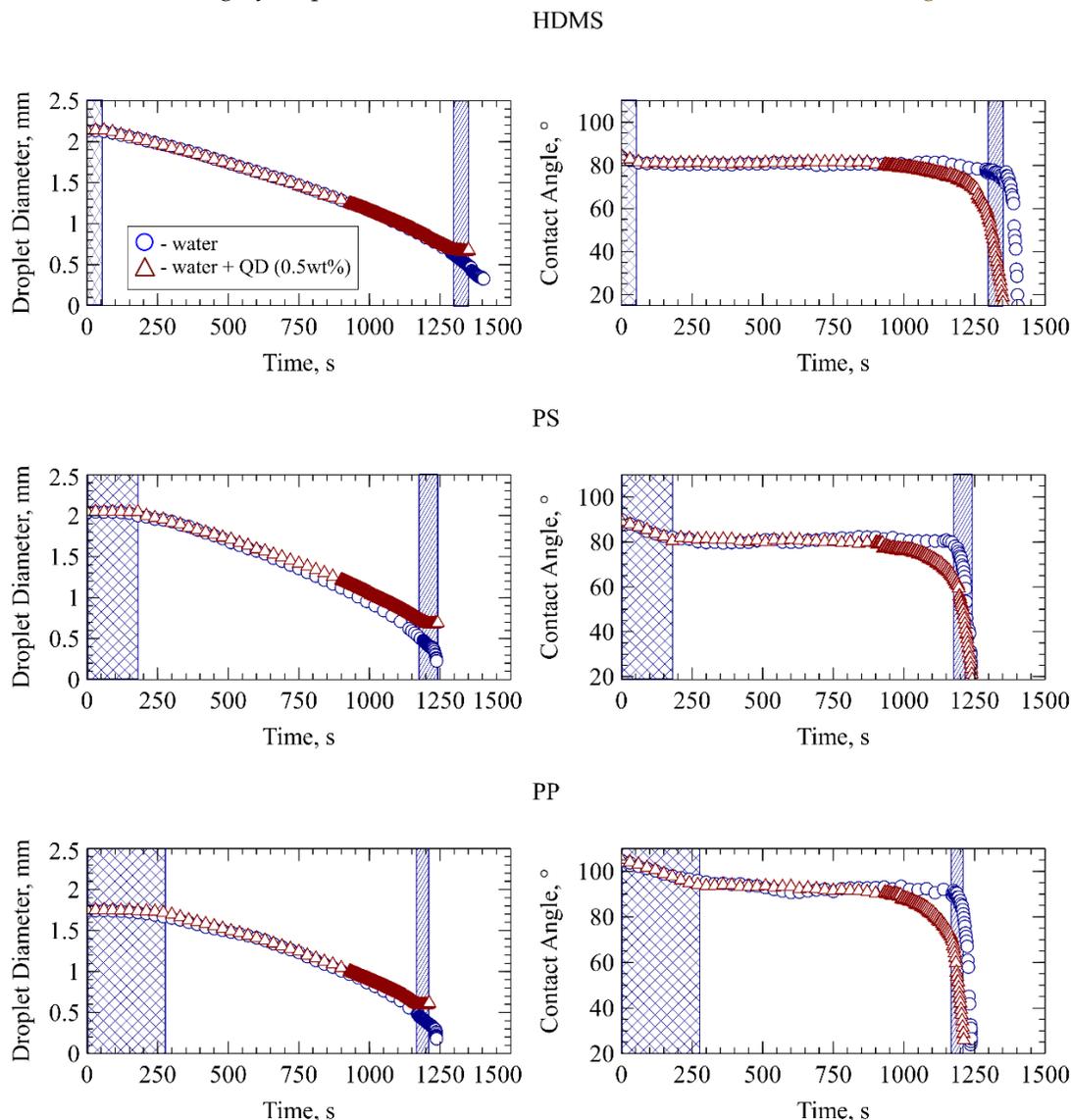

**Figure 2.** Evaporation of pure water and water-based nanosuspensions of QD ($c_{QD}$ = 0.5 wt%) on different surfaces: HDMS ($\theta_w = 84°$), PS ($\theta_w = 90°$) and PP ($\theta_w = 106°$). Crossed areas on each graph correspond to CCR evaporation mode and slant-lined areas correspond to ACCR (Additional Constant Contact Radius) mode caused by the coffee stain formation on the droplet periphery. Volume of droplet deposited is 2.2 µl.

It is well-known that a droplet after deposition on a rigid smooth substrate can go through three different modes during the evaporation. The modes are usually referred to as constant contact radius mode (CCR) if TCL of the droplet is pinned to the substrate and wetted perimeter does not change, the constant contact angle mode (CCA) if TCL of the droplet recedes with unchanging contact angle and mixed evaporation mode, in which droplet loses volume via both wetted perimeter shrinkage and contact angle diminishing.[55-56] Duration of each stage for the droplet of pure liquid depends on the surrounding atmosphere saturation with the vapors of liquid used, temperature, substrate surface energy and surface roughness.[56-58] When nanoparticles are introduced, this duration is additionally dependent on the particle material and size, surface charge and surface screening by ligands.

It can be seen that evaporation of water droplets (blue circle signs on Fig. 2) on each surface examined involves CCR, CCA and mixed mode, respectively. First, the droplet deposited onto the substrate evaporates with unchanging wetted perimeter while contact angle decreases from the initial value $\theta_I$ to contact angle $\theta_R$ which can be referred to as receding contact angle, and afterwards the TCL starts to shrink towards the droplet center.

In surrounding conditions preserved, the duration of pinned stage is mainly affected by two factors: the surface roughness and the surface energy of substrate material. However, despite the fact that on the one hand, CCR prolongs for more rough surfaces (surfaces with high value of contact angle hysteresis) because of the pinning of TCL to surface topographic non-uniformities, and on another hand, the time droplet spent being pinned shortens for more hydrophobic substrates,[55,58] roughness factor tends to be dominating. Fig. 2 shows that the longest CCR mode is observed for the PP-covered glass slide, which has the highest roughness and the highest water contact angle and amid others used, and the shortest – for HDMS-covered surface which is smooth and has the lowest water contact angle.

Nanoparticles as was described before can affect the evaporation dynamics essentially. Despite the fact that the presence of nanoparticles does not affect initial contact angle $\theta_I$ and pendant drop experiments performed revealed that QDs do not modify the surface tension of water, addition of QDs does change the evaporation dynamics. As can be seen from Fig. 2 and Fig. 3(A), the second – CCA – mode undergoes significant changes when nanocrystals are added: while pure water droplet evaporates predominantly preserving its contact angle and has sufficiently sharp transition from CCA mode to mixed mode of evaporation when both contact angle and contact diameter simultaneously decrease, QD-nanofluid droplet deviates from CCA mode much earlier. Additionally, the rate of contact angle decrease is much slower. Note that by this time, crystal-crystal interactions become the main factor affecting the droplet dynamics, since the volume fraction of QDs increases tenfold: from 0.5 wt% up to 5wt% due to evaporation of the base fluid. It is interesting to notice that transition from CCA to mixed mode was examined for hydrophobic surfaces by Park et.al.[59] and the same tendency was revealed: transition to mixed mode was delayed for purified water compared to tap water including impurities.

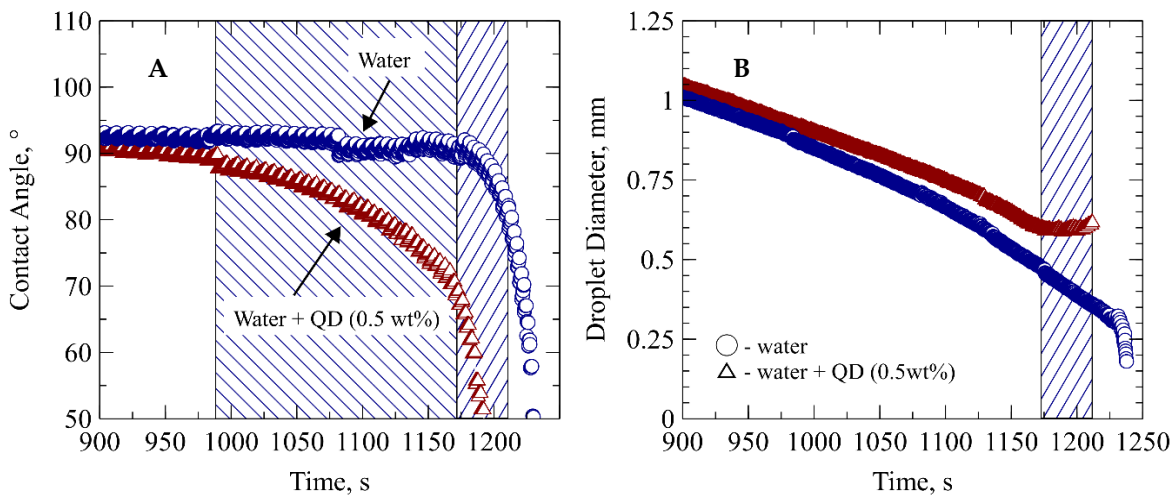

**Figure 3.** Evaporation of pure water and water-based nanosuspensions of QD ($c_{QD} = 0.5$ wt%), ) on PP surface ($\theta_w = 106°$). A: Comparison of the CCA evaporation mode for pure water and nanosuspensions of QDs. Left slant-lined rectangle illustrates mixed mode nanosuspensions, right slant-lined rectangle illustrates ACCR mode for nanosuspension. B: Comparison of the third evaporation modes with and without nanoparticles presence. Slant-lined rectangle corresponds to evaporation in ACCR mode when QDs are added.

It is necessary to highlight, moreover, that the last mode of evaporation in presence of nanoparticles distinguishes from the last mode of evaporation of pure volatile liquid, as follows from Fig. 2 and Fig. 3(B). In addition to the final, mixed, phase of evaporation, inherent to pure liquids, when both wetting perimeter and contact angle simultaneously decrease, QDs-laden nanofluids experience an ACCR (additional constant contact radius) mode. This mode is observed for all surfaces and is shown in Fig. 3(B) by the area filled with slanted lines. The beginning of ACCR occurs when particles agglomerated on the droplet periphery cannot be torn off by the receding wetting front. The presence of ACCR is related to the self-assembly of nanoparticles on the TCL and residual pattern formation. Interestingly, the existence of ACCR mode for drying $SiO_2$-nanoparticle-laden liquids was reported by Nguyen et.al.[60] Authors refer to this mode as "late pinning regime" in contrast with "early pinning" taking place right after droplet deposition. Similar results with late pinning were obtained by Uno et.al.[61] However, in the investigation performed by Nguyen et.al.[60] and by Uno et.al.[61] for SU8 surface ACCR mode replaced mixed mode and no pronounced mixed mode was observed, whilst in our experiments evaporation of nanofluids goes through both mixed and ACCR modes.

Microphotographs of patterns formed after drying of QD-nanofluid are presented in Fig. 4(A). The deposits formed on HDMS and PS look similar. This can be related to their close values of contact angles ($\theta_w = 84°$ and $\theta_w = 90°$ for HDMS and PS, accordingly) and similar roughness values. It is important to notice that in both cases prominent ring-like deposits are observed, notwithstanding the fact that before ACCR mode TCL was pinned only in the beginning of the base fluid evaporation but not throughout the whole process what is one of the usual conditions for the coffee-ring pattern formation. Curiously, recently, the question of using of hydrophobic surfaces to suppress coffee ring effect was discussed by Mampallil et.al.[62]. Sowade et.al. reported that contact angles corresponding to hydrophobic surfaces like HDMS lead to agglomeration of particles in the center of wetted perimeter and no coffee ring effect can be seen.[63]

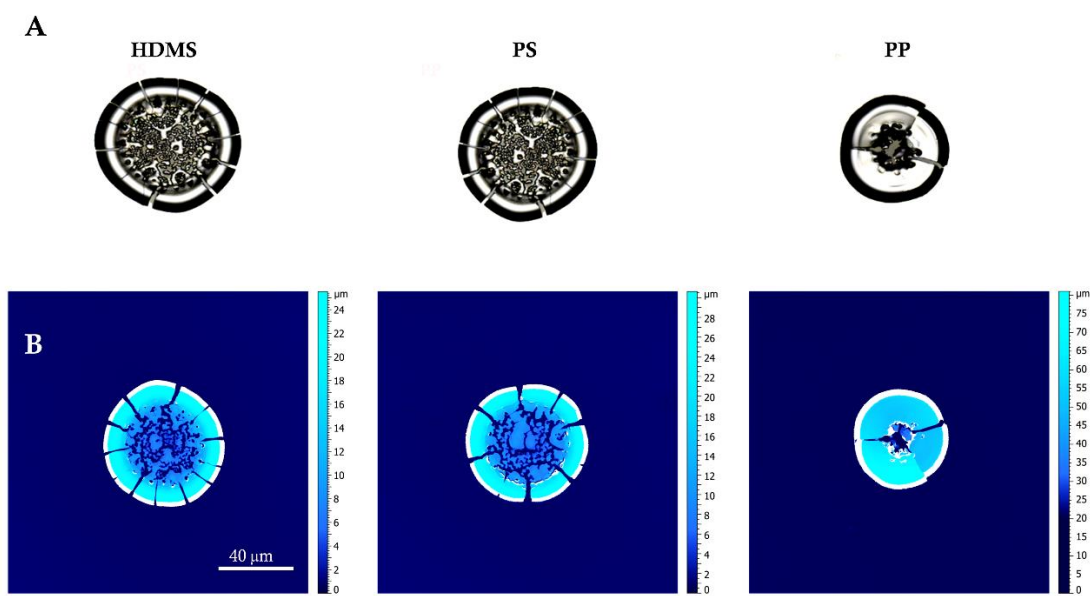

**Figure 4.** Deposits formed evaporation of water-based nanosuspensions of QD (0.5 wt%) on surfaces with different wettability: HDMS ($\theta_w = 84°$), PS ($\theta_w = 90°$) and PP ($\theta_w = 106°$). A: confocal microphotographs show the deposition patterns left after complete drying of nanofluids. B: the topography maps of the coffee stains. The size of each square $1.6\ mm \times 1.6\ mm$. Volume of droplet deposited is 2.2 μl.

From Fig. 2 and Fig. 4, one can observe that no traces of TCL stick-and-slip motion during the receding from the position corresponding to CCR to the one corresponding to ACCR are evident. Despite the different chemistry of the systems under investigation, similar contact line motion was found again by Nguyen et.al and by Uno et.al.[60,61] It was referred to as "Inner Coffee Ring Deposits" by Nguyen et.al.[60] We, however, will save this term for another effect which will be considered in the next subsection.

The QDs-deposit on PP distinguishes from ones obtained for HDMS and PS and has a button shape what is in between conventional coffee-ring structures and dome-like structures. Despite the high roughness of PP surface, the same behavior has been found: the moving TCL carried all nanocrystals when receding toward the droplet center until ACCR mode started and the wetted area during CCR differs from area covered by deposited particles after drying out.

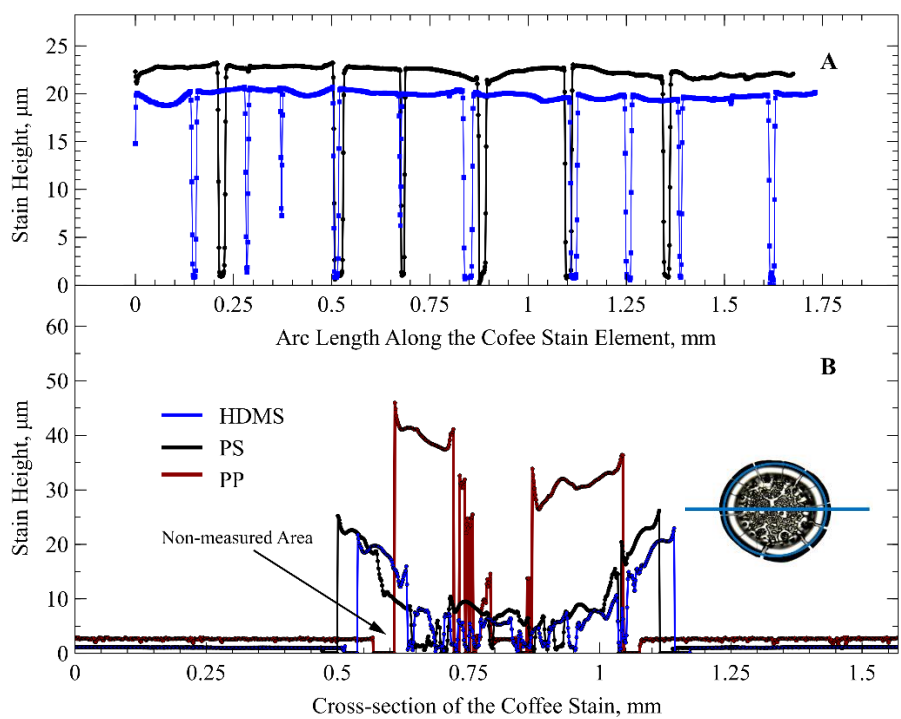

**Figure 5.** Morphology and crack structure of deposition patterns left after drying of water-based nanosuspensions of QDs (0.5 wt%) on surfaces with different wettability: HDMS ($\theta_w = 84°$), PS ($\theta_w = 90°$) and PP ($\theta_w = 106°$) obtained with use of confocal profilometry. A: Cracks distribution along the coffee stain, B: Coffee stain profile. Solid lines on plots are guide to eye. Inner microphotograph of the residual stain and blue lines are to show how parts of deposit was chosen to represent its cross-section height dependence and crack morphology. Zero-set points correspond to out of range non-measured areas (the example of non-measured area is shown in figure by the arrow).

In Fig. 4(B) and Fig. 5(A-B), the topography maps of the stains as well as the droplet profiles and the crack morphologies of the deposition patterns are presented. It is seen that for HDMS and PS, a major part of nanocrystals is collected in the ring. In addition, the inner part of the ring is occupied by particle-formed islands. The appearance of these islands can be related to instabilities in receding wetting front and to adsorption of nanoparticles at solid-liquid interface due to surface forces existing between substrate and particles. We will address the question of surface forces acting in colloidal solutions in following subsection and the question of instabilities in the section next to following. The outer (white-colored) part of all residual structures are out of the scale and, hence, could not be measured to be further transformed to the height map. We suggest that it can probably be related to the blade-sharp deposits' structure in close vicinity to the outer "walls" of the stains. The average height of the coffee rings formed by QDs-nanofluid on HDMS and PS surfaces, measured at approximate middle of the ring as is shown by the blue line in Fig. 5(B), is $h_s = (23 \pm 3)$ $\mu m$. PP-coated substrate does not have developed crack structure (Fig. 4(B)). The average height of the stain on PP surface can be estimated from Fig. 5(B) and is $h_s \approx 40$ $\mu m$. Note that the comprehensive analysis of the droplet profiles and the crack morphologies require further investigation including consideration of influence of the droplet size, nanoparticle concentration, and environmental conditions.

*3.2. Simultaneous Spreading and Drying of QDs-Laden Nanofluid in Presence of Silwet L-77*
*3.2.1. Spreading and Drying of QDs-Laden Nanofluid in Presence of Silwet L-77 at CAC*

In this subsection, we study the effect of introduction of QDs nanocrystals to water-based Silwet L-77 solution on the TCL motion as well as the deposition patterns. In Fig. 6, the droplet diameter and contact angle are presented as functions of time for Silwet L-77 aqueous solution ($c_s$ = CAC) and Silwet L-77 aqueous solution laden with QDs ($c_s = CAC, c_{QD} = 0.5$ wt%). Note that the spreading dynamics is presented only within the timespan of 100 s to be able to compare spreading rate for fluids with and without of nanoparticles. Beyond this time interval, diameter and contact angle determination in case of aqueous solution of Silwet L-77 is hindered since small contact angles lead to high errors in extraction of the goniometric parameters from optical images.

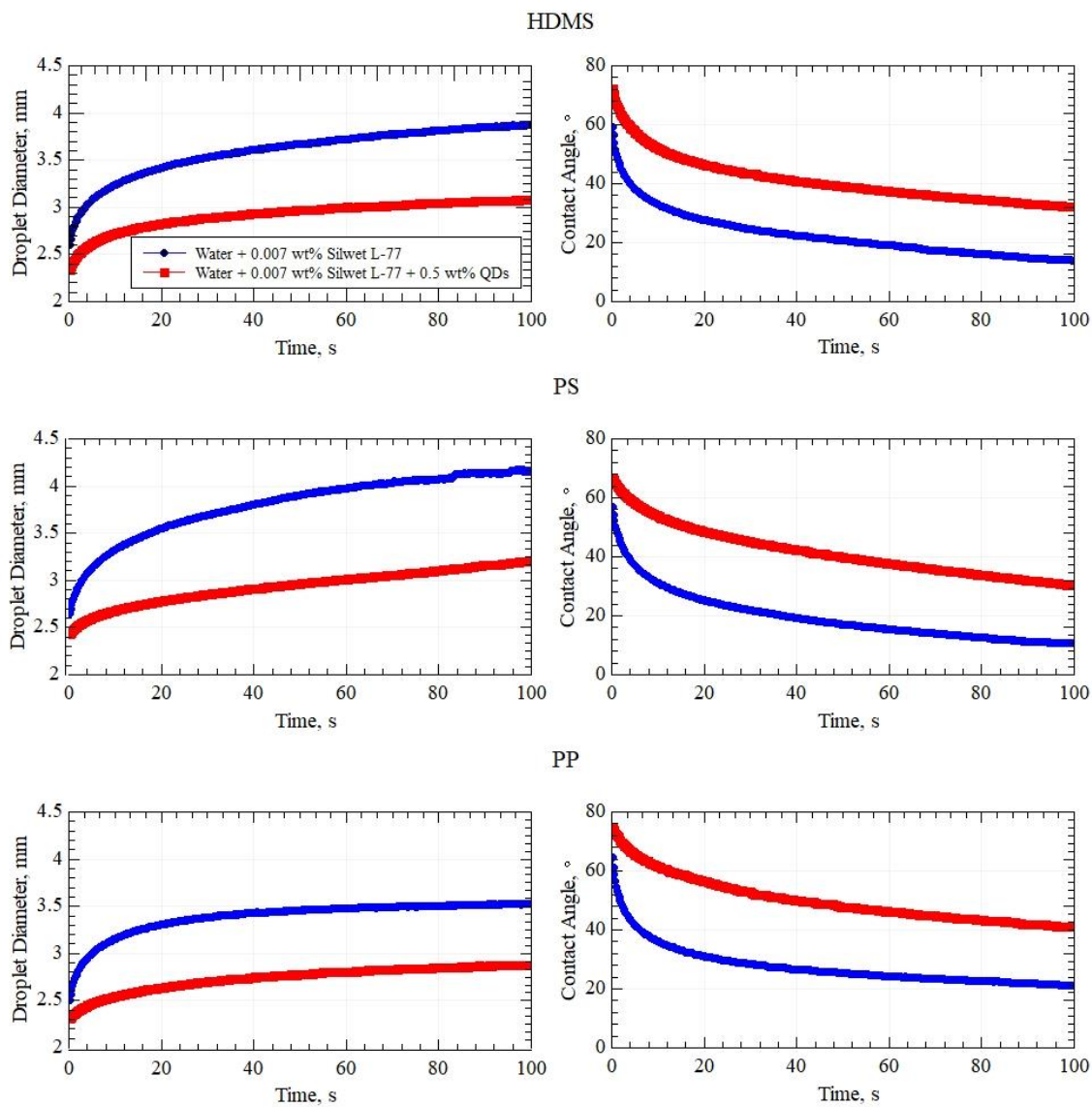

**Figure 6.** Wetting behavior of Silwet L-77 CAC aqueous solutions with ($c_{QD} = 0.5$ wt%, $c_s = CAC < CWC$, denoted by red markers) and without ($c_s = CAC < CWC$, denoted by blue markers) QDs on different hydrophobic surfaces: HDMS ($\theta_w = 84°$), PS ($\theta_w = 90°$) and PP ($\theta_w = 106°$). Volume of droplet deposited is 2.2 μ$l$.

The significant difference in initial contact angles and spreading dynamics between particle-laden and particle-free surfactant solutions can be observed for all substrates (Fig. 6, Table 2). Unexpectedly, nanocrystals-containing surfactant solution shows slower spreading rate on all hydrophobic surfaces (Fig. 6, red curves), which is in contradiction with a number of experimental results obtained before.[10,21-24] The different behavior can be observed since the moment droplets were deposited onto the surfaces: droplets of surfactant-laden nanofluid meet each of examined surface with contact angle higher than surfactant solution. One can also notice that for the HDMS and PP substrates the differences between diameters and contact angles of particle-laden and particle-free surfactant solutions, respectively, are sustained throughout the spreading process.

**Table 2.** Wetting behavior of Silwet L-77 CAC aqueous solutions with and without presence of QDs, $\theta$ – contact angle solution or nanosuspension meets substrate with.

| Substrate<br>Type of Fluid | HDMS | PS | PP |
|---|---|---|---|
| | $\theta, °$ | | |
| 0.007 wt% Silwet L-77<br>(non-nanofluid) | 59±1 | 57±3 | 64±2 |
| 0.007 wt% Silwet L-77 + 0.5 wt% QDs<br>(nanofluid) | 72±2 | 67±2 | 75±1 |

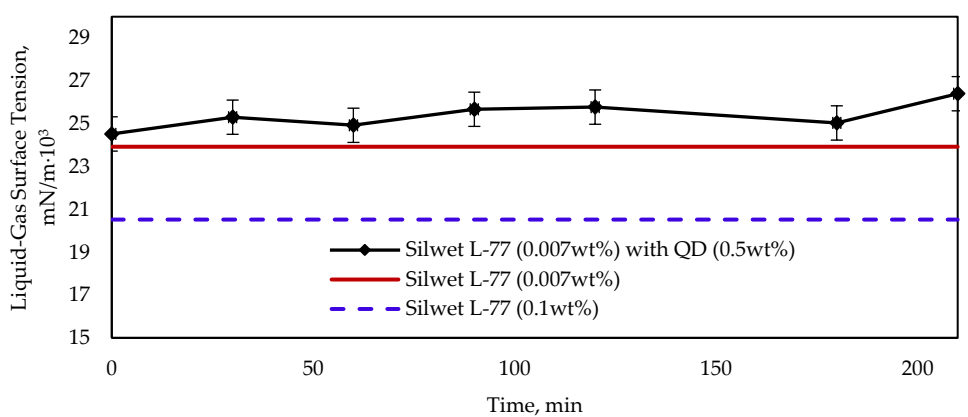

**Figure 7.** Surface tension of QDs nanosuspension with Silwet L-77 ($c_{QD} = 0.5$ wt%, $c_s = 0.007$ wt%), measured by Pendant Drop method. Each surface tension measurement presented was conducted after 1 min of droplet pending. The surface tension of QDs nanosuspension with of Silwet L-77 had been measuring during 3.5 hours. The dotted black line is guide to eye. Solid red and dashed blue lines represent the average surface tension of aqueous solutions of Silwet L-77 at $c_s = 0.007$ wt% and $c_s = 0.1$ wt% > CWC.

Three main factors can be responsible for different behavior of particle-free and particle-laden surfactant solutions: (1) changes in surfaces tension actuated by adsorption of surfactants at interfaces, (2) influence of viscosity and so-called jamming in the close vicinity to the contact line obstructing the TCL motion and (3) ordering of nanoparticles in the wedge which affects Derjaguin's pressure as mentioned in the section 1. In the following, the possible influence of these factors is discussed.

The liquid-gas surface tension $\gamma_{LG}$ of surfactant-laden QD-nanofluid has been measured several times following the preparation and 5 min of ultrasonification (Fig. 7). Indeed, the surface tension of surfactant-laden nanofluid is slightly higher than that of aqueous solution of Silwet L-77 only. However, the curve does not show monotonic increase. Initially, surface tension of surfactant solutions with and without QDs is almost equal. Let us highlight here that despite only averaged results are presented in Fig. 6, single tests in course of set of measurements do not show substantial sensitivity to this minor surface tension increase as well. To estimate the contribution of the increasing $\gamma_{LG}$ to the increase of contact angle, van Oss molecular theory of contact angles[64] can be used assuming that the presence of the adsorbed film does not affect the solid-gas surface energy. Since most hydrophobic polymers are apolar and HDMS-coated surface can be assumed weakly polar,[65] van Oss approach reduces to Berthelot's approach considering only dispersion forces contribution to surface tension,[66,67] which coupled with Young's equation allows to exclude unknown solid-liquid component $\gamma_{SL}$ and yields

$$\cos\theta_2 = \frac{\cos\theta_1 + 1}{\sqrt{\omega}} - 1$$

where $\theta_1$ and $\theta_2$ are contact angles without and with nanoparticles, respectively, and $\omega = \frac{\gamma_{LV_2}}{\gamma_{LV_1}}$ is a magnitude of the surface tension increase calculated as a ratio of liquid-gas surface tensions with and without nanoparticles, respectively. Taking the average contact angles from the Table 2 and average values $\gamma_{LV_2} = 25.4\ mN/m$ and $\gamma_{LV_1} = 23.9\ mN/m$ gives us average contact angle difference $\Delta\theta = \theta_2 - \theta_1 \approx 3°$ for all surfaces, whereas actual increase is more than 10°. It has been checked that the dynamics of spreading does not depend on time elapsed after preparation of suspension within the first 3.5 hours. Taking into account the fact that QDs do not change surface tension of water, one can also conclude that this alteration can be related to the specific surfactant-nanoparticles interactions. Radulovic et.al.[68] showed that Silwet L-77 demonstrates high sensitivity to pH and prone to hydrolysis at low pH. We assume, hence, that slight increase of surface tension of Silwet L-77 aqueous solution can be resulted from presence of acidic –COOH groups of MSA in the solution.

Due to the very small amounts of nanosuspensions, the direct viscosity measurement was not possible. Instead, the viscosity of nanofluid was estimated using Einstein equation.[69] The latter approach is valid since the suspension can be considered diluted (the volume fraction of nanoparticles is approximately 0.08). The estimation resulted in negligible increase (less than 1%) of the viscosity of nanofluids compared to pure water. The local jamming mentioned above also cannot contribute to wetting dynamics in our case because the nanosuspension demonstrated slower TCL velocity since the very beginning of the spreading process when nanoparticles could not be collected near TCL.

Thus, surface tension and viscosity can bring negligibly small changes in wetting behavior of nanosuspension in our case. Therefore, it can be suggested that surface forces acting in a liquid wedge and in a thin precursor film adjoining the droplet are responsible for the wetting dynamics alteration. Note the enormous complexity of the system under consideration: five groups of interactions governing the wetting and self-assembly processes and result the droplet behavior we observe: (1) particle-substrate, (2) particle-particle, (3) particle-surfactant, (4) surfactant-substrate and (5) surfactant-surfactant (aggregate-aggregate) interactions, accordingly. Each of first three interaction types can be again subdivided by two types of DLVO forces: van der Waals force and electrostatic force. Particle-particle interactions in case of stabilized nanoparticles or nanocrystals, as was mentioned before, cannot be completely described in a frame of DLVO theory due to the presence of steric forces. Steric forces are considered to be determined by the ligand properties: type, density, chain length, and charge and affect the interparticle spacing either leading to aggregation of particles or to their repulsion. One also has to keep in mind that ordering of nanoparticles in the confined thin precursor film affects drop dynamics by oscillating structural forces which are reported to be the strongest amid other surface forces.

From a naive mechanical point of view, contact angle increase can only be induced by not disjoining, but conjoining pressure striving to contract the wetting film ahead the droplet in the direction normal to the substrate surface plane. The model for estimation of DLVO interaction potential between nanoparticles shelled with ligand brush is proposed by Yeom et.al.[70] However, the magnitudes and signs of these forces require further investigation.

Let us address the question of deposits left after drying out of surfactant-containing nanosuspensions. Microphotographs of deposition patterns as well as topography maps for nanofluid containing 0.007 wt% of Silwet L-77 and 0.5 wt% of QD on HDMS, PS and PP surfaces are shown in Fig. 8(A-C). When particles are added to surfactant solution of Silwet L-77, prominent ring-shaped deposits are observed. The outer coffee ring size is comparable with diameter reached during the spreading stage. Significantly, not conventional single but double coffee ring appears as a result of simultaneous spreading and evaporation of trisiloxane-containing nanofluid on HDMS and PS surfaces. One can, besides, clearly observe that the structure of the inner ring differs from the outer one: it is thinner and is not equipped with additional craquelure.

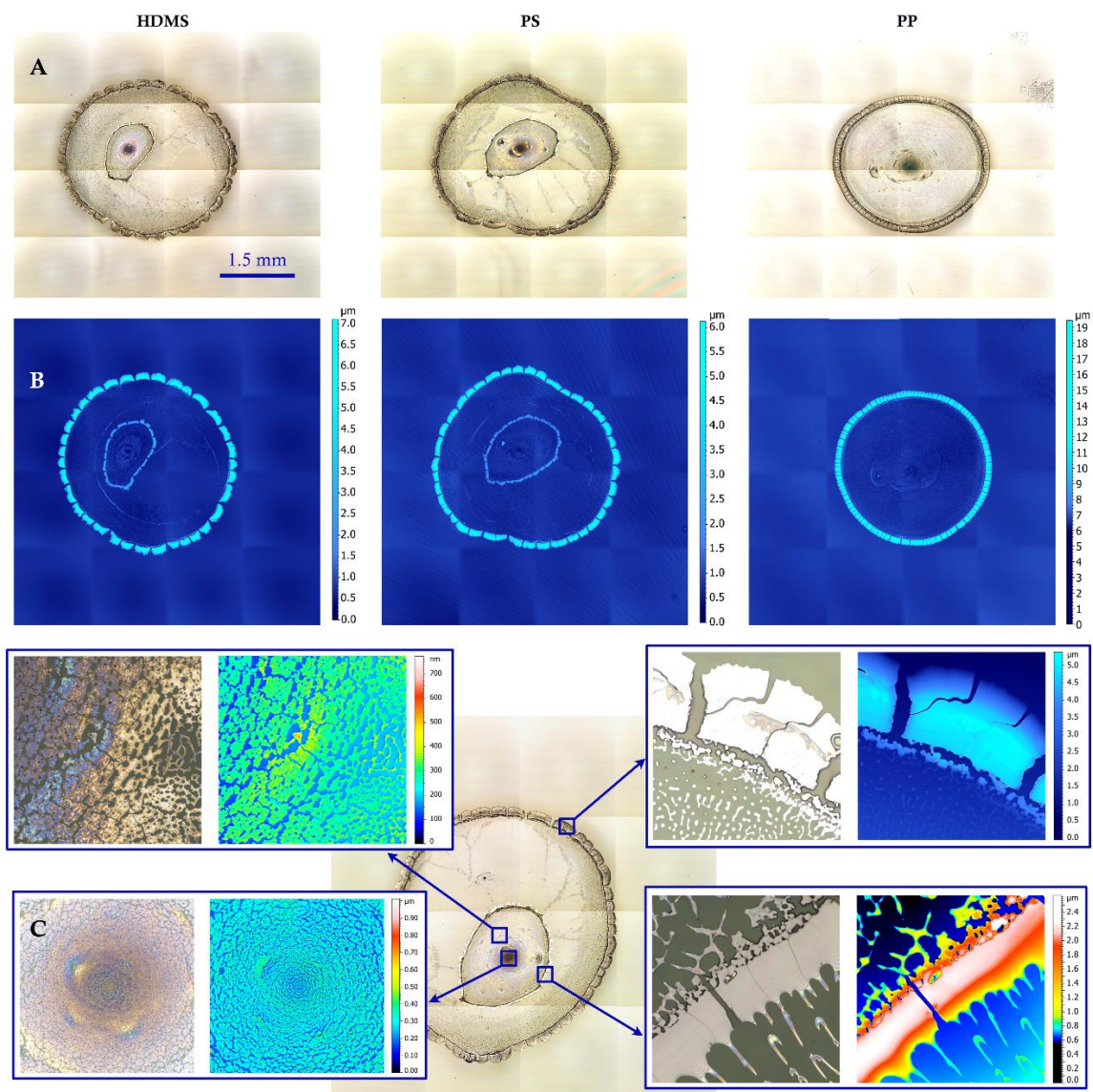

**Figure 8.** Deposition patterns left after spreading and evaporation of Silwet L-77 CAC aqueous solutions with ($c_{QD} = 0.5$ wt%, $c_s = CAC$) and without ($c_s = CAC$) QDs on different hydrophobic surfaces: HDMS ($\theta_w = 84°$), PS ($\theta_w = 90°$) and PP ($\theta_w = 106°$). A: confocal microphotographs of the deposition patterns. B: the topography maps of the deposition patterns. C: microphotographs and morphology maps obtained for the deposition pattern on PS substrate. Size of each inner area scanned for deposition C is 160 µm x 160 µm. Volume of droplet deposited is 2.2 µl.

We suggest that the inner coffee ring is resulting from dewetting process (Fig. 8, C, top and bottom microphotographs in the right corner). In Fig. 9, the formation process of the inner coffee ring on HDMS-covered substrate during one of experimental tests is illustrated. After the droplet is deposited onto the substrate, the spreading begins. After several minutes, droplet has a pancake-like shape resulting from spreading and evaporation and cannot be detected with goniometric set-up ($t = 600\ s$). A few seconds later, an evaporation-induced rupture of the thin film in the vicinity of the boundary of wetted occurs. After that the contraction of TCL toward the initial droplet center leads to an inner smaller droplet formation ($t = 604 -$ 608 $s$). Drying out of the smaller droplet causes appearance of the inner coffee ring. Pronounced dewetting front traces can be observed with use of a picture reconstructed from the height-position matrix (Fig. 9, B).

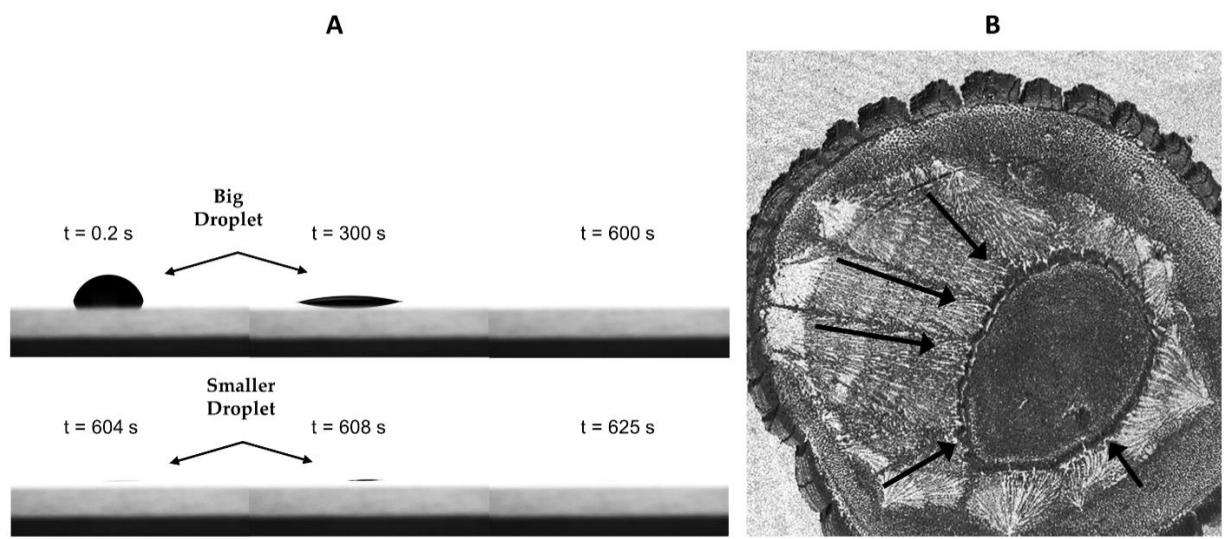

**Figure 9.** A: Time-lapse images of spreading and evaporation of Silwet L-77 CAC aqueous solution with QDs ($c_{QD}$ = 0.5 wt%, $c_{ss} = CAC < CWC$) on HDMS surface. The image sequence shows formation of inner coffee ring. Volume of droplet deposited is 2.1 $\mu l$. B: Topography reconstructed from height-position matrix obtained by confocal profilometer (PS substrate). Receding TCL traces are shown by black arrows.

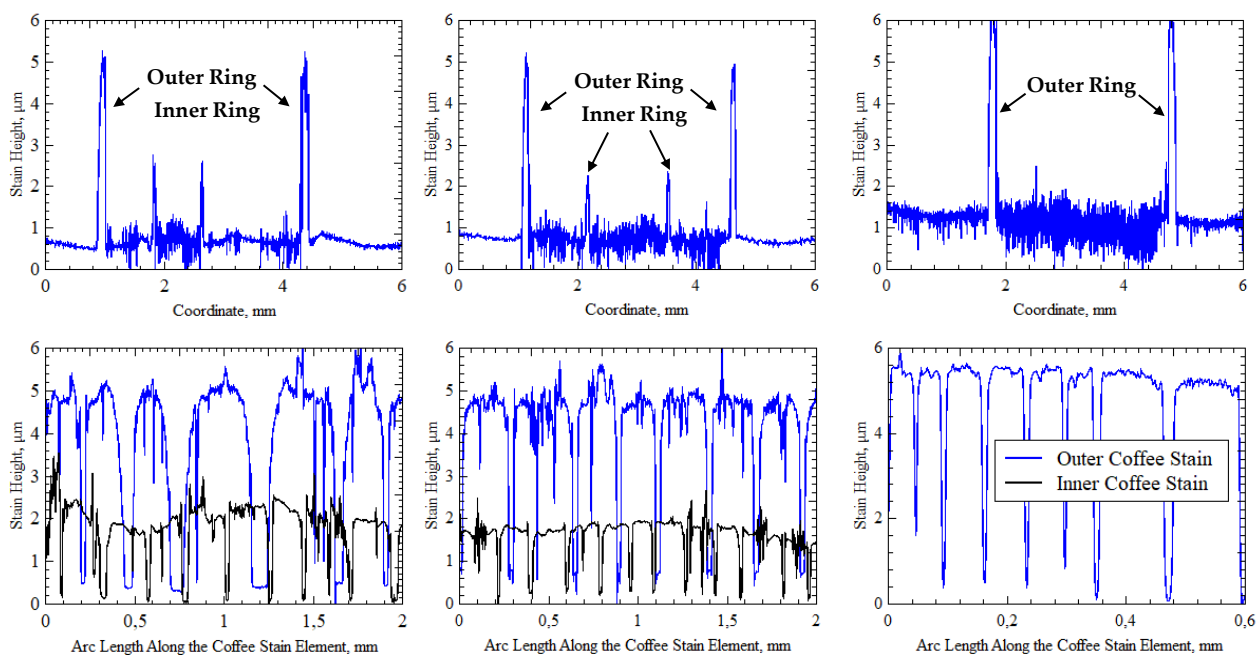

**Figure 10.** Morphology and crack structure of deposition patterns left after drying of trisiloxan-laden nanosuspensions of QDs (0.5 wt%) on surfaces with different hydrophobicity: HDMS ($\theta_w$ = 84°), PS ($\theta_w$ = 90°) and PP ($\theta_w$ = 106°) obtained with use of confocal profilometry. A: Coffee stain profiles. B: Cracks distribution along the coffee stains.

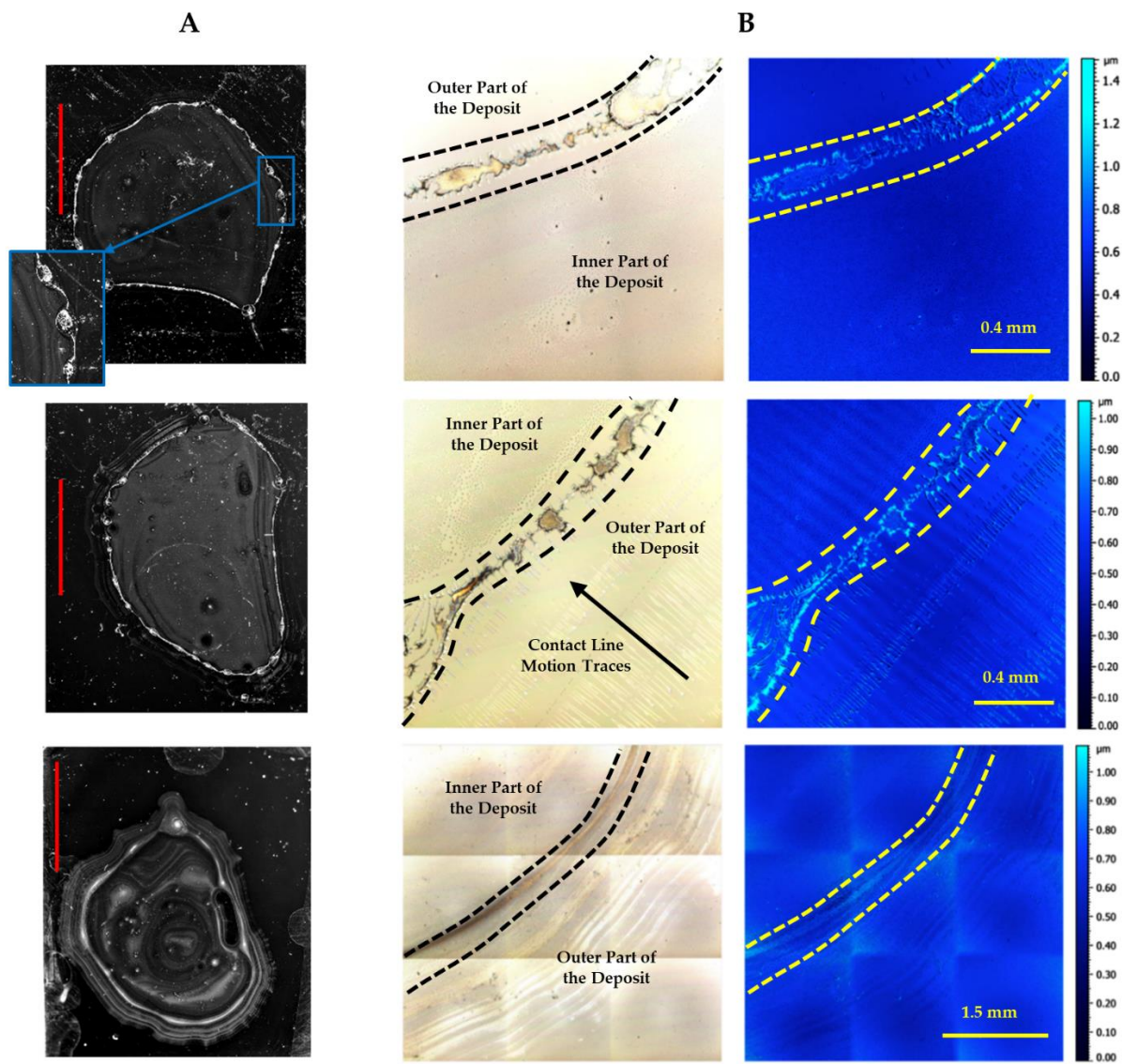

**Figure 11.** Images and morphology of deposition patterns left after drying of nanosuspensions of QD ($c_{QD}$ = 0.5 wt%) laden with Silwet L-77 ($c_{super} = 0.1\ wt\%$) on different hydrophobic surfaces: HDMS ($\theta_w = 84°$), PS ($\theta_w = 90°$) and PP ($\theta_w = 106°$). A: photographs of the patterns left on substrates after drying out of droplets. Red bars correspond to 10 mm scale. Magnified region (blue frame) aims to show non-uniformity of the coffee ring formed. B: confocal microphotographs and topography maps of fragments of contact line regions.

The deposition patterns on PP distinguishes from patterns on PS and HDMS: the inner coffee ring is absent, but a dark area can be observed in the center of the ring center instead. We investigated the ring profiles and fragments of the rings to demonstrate crack morphology of the depositions (Fig. 10, A, B). The height of the ring for trisiloxane-laden droplets is approximately fivefold smaller than for water nanosuspensions on HDMS and PP and is approximately tenfold less on PP surfaces. We also observe that addition of surfactant makes relatively smooth surface of cracks more peaky and serrated on HDMS and PS, whereas on PP it has more regular shape. The height of the inner coffee ring is almost threefold lower than outer conventional coffee ring.

*3.2.2. Superspreading and Drying of QDs-Laden Nanofluid*

If the surfactant concentration is increased above CWC, superspreading is observed. Superspreading is referred to as a phenomenon in which a sessile droplet spreads out over the substrate until zero contact angle is reached. Below, we present the first results on superspreading in the presence of nanoparticles. All superspreading experiments were conducted at the same temperature as experiments with surfactant concentration below CWC, which are described in the previous subsection, but higher humidity: T = (23±2)°C, RH = (45±5)%. Lin et.al.[71] showed that the superspreading is highly sensitive to concentration and that the maximum spreading rate for droplet of Silwet L-77 on hydrophobic substrate is observed at $c_{super} = 0.1\ wt\%$. The same concentration corresponding to maximum wetted perimeter is reported by Venzmer.[2] We have chosen this concentration for superspreading experiments. The surface tension of Silwet L-77 aqueous solutions measured by pendant drop method reaches the equilibrium value instantly and is equal to $\gamma_{LG\ super} = (20.8 \pm 0.2)\ mN/m$ (Fig. 7).

Superspreading effect in absence as well as in presence of QDs was observed for all substrates used (HDMS, PS, PP). However, since a droplet containing superspreader surfactant at $c_{ss}$ spreads out within 2-3 s becoming undistinguishable from the side-view, only the images of deposit patterns are presented here (Fig. 11, A-B). Drying patterns on HDMS and PS indicate pronounced coffee ring effect (average ring height ~ 1 $\mu m$). Interestingly, rings have non-uniform thickness and equipped with "knots" (Fig. 11, A, magnified region). On more hydrophobic PP-covered surface, nebula-like stain is observed. The non-uniform shape of final deposition patterns can be attributed to the imperfections of the substrates. The contact line motion is known to be strongly sensitive to any surface defects if the contact angle is low.

Another interesting observation is related to highly pronounced spike dewetting patterns left when TCL was receding (Fig. 11, B) which allows to suggest that the diameter reached during the spreading stage is bigger than the one corresponding to the coffee ring. When spreading is finished and droplet has a pancake-shape, the evaporation process comes into play forcing the TCL to shrink toward its center and as a result spiked pattern precedes the coffee ring formed. Similar spike-producing dewetting was observed for evaporation of water surfactant solutions of DTAB by Bernardes et.al.[72] and, what is more interesting, for evaporation of toluene containing CdSe/ZnS QDs by Xu et.al.[73] Spike dewetting traceries are frequently attributed to fingering in the receding wetting front. Xu et.al.[73] reported that development of fingering instabilities depends on the rate of the TCL: slow dewetting front velocity induces instable behavior of the TCL. We suggest that in our case receding of the dewetting front motion decelerates due to the nanoparticle accumulation in the close vicinity of the TCL, which leads to the formation of the nucleation sites and to their further extension to spikes. Structure of patterns is most probably related to stick-and-slip motion of the contact line in the course of evaporation.

For PP-coated surfaces, we did not observe the formation of spike structures. As is seen from Fig. 11 (B), the TCL moves in stick-and-slip regime.

## 4. Conclusions

In this paper, experimental investigation of drying and spreading dynamics of complex nanofluids containing quantum dot nanocrystals and trisiloxane superspreader surfactant Silwet L-77 on hydrophobic surfaces has been performed. It was revealed that drying of surfactant-free quantum dot nanofluids in contrast to pure liquids undergoes four evaporation modes: constant contact radius, constant contact angle, mixed mode, when both wetted perimeter and contact angle decrease, and additional constant contact radius mode when contact angle decreases whilst triple contact line is pinned. After drying, nanosuspension droplets form cracked ring-shaped deposition patterns on HDMS and PS but assembled into button-like pattern the PP substrate, which is characterized by higher static contact angle and by a higher roughness. For all substrates, coffee ring areas are smaller than area wetted by droplets after deposition.

Addition of trisiloxane surfactant promotes spreading of droplets over hydrophobic surfaces. We found out that the spreading rate after addition of nanoparticles was slower than for water surfactant solutions what is in contrast with a number of works reporting on the enhanced spreading of nanofluids induced by the gradient of structural disjoining pressure. However, we do not relate this deceleration with surface tension and viscosity effects. We suggest that observed wetting dynamics can be resulted from complex interplay of surfaces forces (DLVO forces, steric, structural forces). Deposition patterns when superspreader surfactant is added essentially distinguish from ones obtained for water-based nanocolloidal suspensions – one can observe double coffee ring effect. The formation of the distinct inner coffee ring, as we suggest, begins when at the late stage of evaporation when droplet accepts a thin pancake and is followed by the film rupture and the movement of the wetted perimeter toward the droplet center. Such film contraction leads to formation of a new droplet, drying out of which gives the inner coffee ring.

For the first time, the superspreading effect in presence of nanoparticles has been observed. Despite the formation of coffee rings on all substrates, they have different morphology. Particularly, the knot-like structures are incorporated into the ring on HDMS and PS surfaces. We found out that the deposit formation process in the course of superspreading distinguishes from the deposit formation process for lower, CAC, concentration: coffee stains corresponding to superspreading concentration form not in the course of the TCL advancing but when it recedes as was for surfactant-free nanofluids. We suggest that the aligned spikes in the vicinity of the ring are caused by the fingering instability of dewetting front.

**Acknowledgments:** The authors are pleased to acknowledge the financial support from the DAAD (Deutscher Akademischer Austauschdienst), and from Deutsche Forschungsgemeinschaft (DFG, German Research Foundation) – Project-ID 265191195 – SFB 1194, subproject A04. Authors would like to thank Dr. Viktor Fliagin (Russia, University of Tyumen, Photonics and Microfluidics Laboratory) for development of the droplet shape analyzing software. Authors also would like to thank Institute of Printing, Science and Technology (TU Darmstadt, Germany) for the help in organizing of experiments. Particularly, authors would like to express a gratitude to Thorsten Bitsch.

**Conflicts of Interest:** The authors declare no conflict of interest.